\documentclass[12pt,preprint]{aastex}

\newcommand{\cii}{{\rm C}\,{\sc ii}}
\newcommand{\ci}{{\rm C}\,{\sc i}}
\newcommand{\mum}{\,\mu{\rm m}}

\shorttitle{CO J=6--5, 4--3, 3--2, and HCN 4--3 emission in Mrk\,231}
\shortauthors{Papadopoulos, Isaak, \& van der Werf}
\begin{document}

\title{First CO J=6--5, 4--3 detections in local ULIRGs:  
the dense gas in Mrk\,231, and its cooling budget}

\author{Padeli \ P.\ Papadopoulos}
\affil{Institut f\"ur Astronomie, ETH Zurich, 8093 Z\"urich, Switzerland}
\email{papadop@phys.ethz.ch}

\author{Kate G. Isaak}
\affil{School of Physics and Astronomy, Cardiff University, Cardiff CF24 3AA, UK}
\email{kate.isaak@astro.cf.ac.uk}

\and 

\author{Paul P.~van der Werf}
\affil{Leiden Observatory, Leiden University,  P.O.~Box~9513, NL-2300 RA Leiden, The Netherlands}
\email{pvdwerf@strw.leidenuniv.nl}

\begin{abstract}

We report on detections of the high-excitation CO J=6--5, J=4--3 lines
in Mrk\,231, a prototypical Ultra Luminous Infrared Galaxy (ULIRG) and
Seyfert~1 QSO.  These observations are combined with CO J=3--2, HCN
J=4--3 (this work), and CO J=2--1, J=1--0, $ ^{13}$CO J=2--1, HCN
J=1--0 measurements taken from the literature to provide better
constraints on the properties of the molecular gas in an extreme
starburst/QSO in the local Universe.  We find that the CO J=4--3 and
J=6--5 transitions trace a different gas phase from that dominating
the lower three CO transitions, with $\rm n(H_2)\sim (1-3)\times 10^4
\ cm^{-3}$ and $\rm T_k\sim (40-70)\, K $. This phase is responsible
for the luminous HCN emission, and contains most of the H$_2$ gas mass
of this galaxy.  The total CO line cooling emanating from this dense
phase is found similar to that of the [\cii] line at $158\,\mu m$,
{\it suggesting a very different thermal balance to that seen in lower
  IR-luminosity galaxies}, and one likely dominated by dense
photon-dominated regions.  Our dense ``sampling'' of the CO rotational
ladder and the HCN lines enables us to produce well-constrained
Spectral Line Energy Distributions (SLEDs) for the dense molecular gas
in Mrk\,231 and compare them to those of high redshift starbursts,
many of which have SLEDs that may be affected by strong
lensing. Finally, we use our local molecular line excitation template
to assess the capabilities of future cm and mm/sub-mm arrays in
detecting CO and HCN transitions in similar systems throughout the
local and distant Universe.

\end{abstract}


\section{Introduction}

Luminous and Ultraluminous Infrared Galaxies ((U)LIRGs) represent a
truly remarkable class of objects. With far-infrared luminosities of
$\rm L_{FIR}\geq ~10^{11-12} L_{\odot }$, these galaxies emit a
significant fraction of their total bolometric luminosity at infrared
wavelengths (Soifer et al.  1986; Sanders 1989).  Although these
rapidly evolving and often merging objects are frequently hosts of both
powerful starbursts and AGN (Sanders \& Mirabel 1996), there is
increasing observational evidence to suggest that most of this IR and
FIR emission is produced by extreme star formation events (Genzel et
al.  1998, Downes \& Solomon 1998).  Observations suggest that the
evolutionary track followed by ULIRGs once their molecular gas supply
is exhausted will likely be towards field elliptical/lenticular
galaxies of moderate mass (Genzel et al.  2001; Tacconi et al.  2002),
rather than towards hosts of optically bright QSOs as originaly
suggested by Sanders et al.~(1988).

Several studies of the molecular gas have been made since its
ubiquitous presence in (U)LIRGs was first established (Tinney et
al.~1990; Sanders, Scoville \& Soifer 1991; Solomon et al.~1997; see
Sanders 1997 for a review).  Intense far-UV radiation and strong tidal
fields in these merging/starburst objects are likely to be responsible
for the differing physical properties of the molecular gas relative to
those prevailing in lower-power starbursts (Aalto et al.\ 1991, 1995;
Casoli, Dupraz \& Combes 1991), while interferometer maps reveal
compact ($\sim (0.5-1)$\,kpc) CO-emitting regions (Bryant 1997; Downes
\& Solomon 1998; Bryant \& Scoville 1998).  A large fraction of the
total molecular gas mass in these objects is in a warm, dense gas
phase (Solomon, Radford, \& Downes 1990; Solomon, Downes, \& Radford
1992a), and the same type of gas may be also responsible for the
bright high-excitation CO lines detected in several dust-enshrouded
and more luminous starbursts and QSOs ($\rm L_{FIR}\sim
10^{13}\ L_{\odot }$) at high redshifts over the last decade (see
Solomon \& Vanden Bout 2005 for a review).

Detections of high-excitation CO transitions such as J=4--3 and J=6--5
in local galaxies were first made in the early 1990s (e.g., Harris et
al.~1991, Wild et al.~1992, G\"usten et al.~1993).  Poor atmospheric
transmission under all but the very best observing conditions has
prevented systematic surveys of these transitions in galaxies other
than in local, compact ($\sim 500$ pc) and low luminosity ($\rm
L_{FIR}\sim 4\times 10^{10}\, L_{\odot }$) starbursts such as M\,82
(White et al.  1994; Mao et al.  2000, Wild et al.  1993, Ward et al.
2003), and the centers of a few nearby spirals (Nieten et al.  1999).
The lack of large-format array receivers at these frequencies makes
imaging of high excitation CO transitions difficult except with
special instrumentation (Fixsen, Bennett, \& Mather 1999; Kim et al.
2002).  This is unfortunate since, when combined with lower-J CO and $
^{13}$CO transitions, CO J=4--3, J=6--5 line emission with $\rm
E_4/k_B\sim 55$ K, $\rm E_6/k_B\sim 116~K$ and critical densities of
$\rm n_{43}\sim 1.9\times 10^4\, cm^{-3}$ and $\rm n_{65}\sim 6\times
10^4\, cm^{-3}$, can provide excellent diagnostics of the excitation
conditions of the molecular gas.

To fill this gap, we initiated a multi-transition CO, $ ^{13}$CO and
HCN survey of $\sim 30$ LIRGs with which to undertake a detailed study
of their molecular gas component, with particular emphasis on the warm
and dense phase fueling their prodigious star formation rates.  In
this paper we report on the principal result of the ULIRG/QSO
Mrk\,231, the first galaxy in our sample for which multi-transition
observations have been completed.  We present an assesssment of the
physical conditions of its molecular gas reservoir based on an LVG
analysis, which we then use to compute the cooling budget of the
reservoir and to construct CO and HCN Spectral Line Energy
Distributions (hereafter SLEDs).  We compare the Mrk231 SLEDs to the
typically poorly constrained SLEDs of high-z starbursts, and use them
to assess the capabilities of both future cm/mm interferometers and
ESA's high frequency spaceborne observatory Herschel in detecting
these important line diagnostics in starbursts in the distant
Universe.  Throughout this work we assume a flat $\Lambda $-dominated
cosmology with $\rm H_0=71\,$km\,s$^{-1}$\,Mpc$^{-1}$ and $\Omega_{\rm
  m}=0.27$.

\section{Observations and results}

The 15-meter James Clerk Maxwell Telescope (JCMT)\footnote{The James
Clerk Maxwell telescope is operated by the Joint Astronomy
Centre on behalf of the Science and Technology Facilities Council of
the United Kingdom, the Netherlands Organisation for Scientific Research
and the National Research Council of Canada.}
on  Mauna Kea  in Hawaii  was  used for
observations of  the CO  J=6--5, CO J=4--3,  CO J=3--2 and  HCN J=4--3
transitions in  the archetypal ULIRG/QSO  Mrk\,231. These observations
are part of  our large and ongoing CO, $ ^{13}$CO  and HCN line survey
of  30 LIRGs  ($\rm L_{IR}\geq  10^{11}\, L_{\odot}$)  which  is being
undertaken  using   the  JCMT  and   the  IRAM  30-m   telescope  (see
Papadopoulos et al. 2007 for a description).

\subsection{The CO J=6--5, 4--3 observations}

Receiver W at D-band (620-710\,GHz) was used on February 20, 2005 in
SSB mode to observe the CO J=6--5 line ( $\rm \nu_{rest}=691.473\,
GHz$) under excellent atmospheric conditions ($\tau _{225}\la 0.035$),
with typical system temperatures of $\rm T_{sys}\sim (4500-5200)\, K$
(including atmospheric absorption).  The Digital Autocorrelation
Spectrometer (DAS) was deployed in wideband-mode (1.8\,GHz), with a
resulting instantaneous bandwidth of $\rm \sim 780\, km\, s^{-1} $
which was more than sufficient to cover the full width of the CO line
($\rm FWZI \sim 400\, km\, s^{-1}$).  Rapid beam switching with a chop
frequency of 1-2\,Hz and a beam throw of 60$''$ (in Az) yielded flat
baselines after a total of 2 hours of integration (on+off).  Pointing
at D-band is complicated by the paucity of suitably bright, compact
pointing sources. Differential pointing was therefore employed using a
combination of B-band (330-360\,GHz, B3 receiver) and D-band
measurements, with the relative pointing offsets of the two different
receivers determined prior to the observing run.  Initial pointing and
focus checks were made using Jupiter which, at the time of the
observations, had a semi-diameter of $\sim 20''$ -- we note that
pointing using sources that are extended relative to the primary beam
has been shown to work well at these high frequencies.  Pointing
checks with B3 were made prior to, and during our observations, and
were applied to the telescope pointing model for D-band, resulting in
a typical rms scatter in the latter of $\sim 2''-3''$.  The CO J=4--3
line ($\rm \nu _{rest}=461.0407\, GHz$) was observed with receiver W
tuned to SSB mode in C-band (430-510\,GHz) on April 23 2005.  Dry
conditions ($\tau _{225}\sim 0.035$) yielded typical system
temperatures of $\rm T_{sys}\sim 1900\, K$. A DAS bandwidth of
920\,MHz ($\rm \sim 600\, km\, s^{-1}$) was used, which was sufficient
to cover the line with ample margin for baseline subtraction.  Beam
switching at 1-2\,Hz with a beam throw of 30$''$ (Az) produced
excellent flat baselines after a total of 40 mins (on+off) of
observations. Pointing checks were again differential, with an rms
uncertainty of $\sim 3''$.

Estimates of the aperture efficiencies at both frequencies were made
from repeated observations of Mars, resulting in a value of $\rm \eta
_a (461\, GHz)=0.36$ (for $\rm \theta _{HPBW}=11''$) which is
consistent within $\sim 10\%$ with values reported by others for the
same observing
period.\footnote{http://www.jach.hawaii.edu/JCMT/spectral\_line/Standards/eff\_web.html}
The scarcity of $\rm \eta _a$ measurements at 691\,GHz did not allow
such comparisons and so we adopted the mean value, $\rm \eta _a(691\,
GHz)=0.18$, of our two measurements ($\rm \theta _{HPBW}\sim 9''$ from
a Mars beam map).  This value has a larger uncertainty -- of around
30\% -- due to larger calibration uncertainties as well as the effects
of thermal distortions/mechanical deformations of the dish affecting
high-frequency observations even on prime sub-mm telescopes such as
the JCMT.

\subsection{The CO J=3--2 and HCN J=4--3 observations}

The CO J=3--2 (345.7960 GHz) and HCN J=4--3 (354.5054 GHz)
observations were made on July 10th 1999 (CO) and January 26th 2005,
January 17th 2006 (HCN), with receiver B3 tuned in SSB mode with
effective system temperatures of $\rm T_{sys}=670\, K$ and $\rm
T_{sys}=(390-470)\, K$ respectively.  A DAS configuration with $\sim
1.8$\,GHz bandwidth was used for CO J=3--2 ($\rm \sim 1556\,
km\,s^{-1}$), whilst a narrower-band 920\,MHz/dual-channel mode was
used for HCN J=4--3 ($\rm \sim 776\, km\, s^{-1}$) for increased
sensitivity. Beam switching at 1 Hz and a chop throw of 60$''$~(Az)
was employed in both cases, giving flat baselines for total
integration times of 20\,min (CO) and 4 hours (HCN).  A number of
aperture efficiency measurements were made using Mars, yielding
$\langle \eta _a \rangle \sim 0.515\pm0.08$ (for $\rm \theta
_{HPBW}=14''$). Frequent pointing checks were made by observing strong
sources in both continuum and spectral line mode, yielding an rms
uncertainty of $\sim 3''$ for the pointing model residuals.  Finally,
observations of strong spectral line standards such as IRC\,10216,
OMC1, W75N and W3(OH) were used to verify the amplitude calibration,
to estimate calibration uncertainties ($\sim 15\%$), and to monitor
the overall performance of the~telescope.

\subsection{Data reduction, results}

All data were reduced using the JCMT spectral line reduction package
SPECX.  Zero-order baselines were removed and spectra inspected
individually, prior to being co-added to produce the final spectra
shown in Figs.~1 and 2, where an excellent line profile agreement,
with a $\rm FWZI\sim 400\,km\,s^{-1}$, is evident for all the
transitions observed.  The velocity-integrated line flux densities
were estimated from these spectra using

\begin{equation}
\rm S_{line}=\int  _{\Delta V}  S_{\nu } dV  = \frac{8 k_B}{\eta  _a \pi
D^2} K_c (x)\int  _{\Delta V} T^* _A dV=  \frac{15.6 (Jy/K)}{\eta _a} K_c(x)
\int _{\Delta V} T^* _A dV,
\end{equation}

\noindent
where the term $\rm K_c (x) =x^2/(1-e^{-x^2})$, with $\rm x=\theta
_s/(1.2\theta _{HPBW})$ and $\theta _s$=source diameter, accounts for
the geometric coupling of the gaussian component of the beam with a
finite-sized, disk-like source.  For Mrk\,231 we have used a value of
$\theta _s\sim 3''$ (likely to be even smaller for the high-J CO
transitions and the HCN emission), obtained from interferometric
observations of CO J=1--0 (Downes \& Solomon 1998), yielding an upper
limit of $\rm K_c (x)\sim 1.04$ in the case of the CO J=6--5 line.
The line fluxes obtained from observations presented in this paper,
those extracted from observations reported in the literature and 
brightness temperature line ratios are listed in Tables 1 and
2.

\subsubsection{CO J=1--0 line flux: single dish versus interferometers}

It can be seen from Table 1 that, with the exception of a single
measurement by Kr\"ugel et al., all CO J=1--0 single dish measurements
are in good agreement with oneanother, but that they are higher than
those obtained using interferometers (which are, in turn, good
agreement among themselves).  The difference between the average of
all the single dish data $\rm \langle S_{10}\rangle _{SD}=(88\pm
9)\,Jy\,km\,s^{-1}$ and that of the two interferometric measurements
$\rm \langle S_{10}\rangle _{INT}=(65\pm 7)\, Jy\,km\,s^{-1}$ is
significant at a $\sim 2\sigma$ level.  If confirmed at a higher
significance level, this discrepancy may result from the presence of a
colder and more extended gas phase that is missed in typical
interferometer maps because of its low brightness (and the resulting
low S/N per beam), rather than the lack of short baselines.  The dust
content of such a gas phase, present even in some luminous
starburst/QSOs, has been detected via its sub-mm continuum emission
(e.g.  Papadopoulos \& Seaquist 1999a), where dust with $\rm
T_{dust}\sim 15\,K$, concomitant with HI and low surface brightness CO
J=1--0 emission has been found.  Sub-mm continuum imaging and
multi-transition CO observations of LIRGs have shown that unless one
has sufficient angular resolution to separate the warm, star-forming 
gas and dust from the typically, more extended non star-forming phase, 
the {\it global} dust continuum and CO line ratios will
be dominated by the warm gas (e.g., Papadopoulos \& Allen 2000).
In this paper we assume that such an extended phase of
non star-forming molecular gas is not present in Mrk\,231.

\section{The state of the molecular gas in Mrk\,231}

The combination of the large number of CO and HCN line detections for
the ULIRG/QSO Mrk\,231 (Table~2) and the upper limit on the $\rm
^{12}CO/ ^{13}CO$ (J=2--1) line ratio of $\rm R_{21}\ga 37$ (Glenn \&
Hunter 2001) offers an excellent opportunity to place strong
constraints on the state of its molecular gas, with emphasis on the
dense gas phase. To do this, we used a Large Velocity Gradient (LVG)
code, based on work by Richardson 1985, that searches a large grid of
$\rm (n, T_k, \Lambda _x)$ values, where $\rm \Lambda _x =
r_x/(dV/dr)$ ($\rm r_x=[X/H_2]$, $\rm dV/dr$: cloud velocity gradient,
X: the molecule used) and locates the $\rm \chi^2=\sum _{i} 1/\sigma
^2 _i [R_i-R_{obs,i}]^2$ minima, where $\rm R_{obs,i}$ and $\rm R_{i}$
are observed and model line ratios, and $\rm \sigma _i$ is the
measurement uncertainty.  A parameter space of $\rm n=(10^2-10^9)
cm^{-3}$, $\rm T_{k}=(15-150)\, K$, and an $\rm \Lambda _x$ range
corresponding to $\rm K_{vir}\sim 0.05-500$ (for the standard CO and
HCN abundances) was searched, where

\begin{equation}
\rm K_{vir}=\frac{\left(dV/dr\right)_{obs}}{\left(dV/dr\right)_{virial}}\sim 
1.54\frac{r_{x}}{\sqrt{\alpha}\Lambda _{x}}\left(\frac{n(H_2)}{10^3\,
 cm^{-3}}\right)^{-1/2},
\end{equation}

\noindent
(e.g.  Papadopoulos \& Seaquist 1999b; Goldsmith 2001) indicates
whether a virialized gas phase ($\rm K_{vir}\sim 1$ within factors of
2-3), or a non-virialized one ($\rm K_{vir}\gg 1$) is responsible for
the molecular line emission ($\alpha \sim 1-2.5$ depending on the
assumed cloud density profile, Bryant \& Scoville 1996).  
Strictly speaking, values of $\rm
K_{vir}\ll 1$ are not physical (i.e. gas motions cannot be slower than
those dictated by self-gravity), however they can be used to
indicate the possibility of enhanced molecular abundances instead (so
that the revised $\rm K_{vir}$ can reach at least $\sim 1$).
Additional constraints to the modelling are provided from estimates of
the dust temperature -- $\rm T_{dust}(Mrk\,231) \sim (47-54)\,K$
(Glenn \& Hunter 2001; Gao \& Solomon 2004a) -- and the assumption
that photoelectric and/or turbulent gas heating (and its cooling via
atomic/molecular lines rather than continuum emission) results in $\rm
T_{k}\geq T_{dust}$ (e.g.  Wilson et al.  1982; Tielens \& Hollenbach
1999). The chosen range for $\rm T_k$ therefore encompasses that
expected for the molecular gas in Mrk\,231, with the temperature of
C$^+$-cooled, Cold Neutral Medium HI gas setting its upper~limit.

The results of single-phase modeling make it clear that {\it the CO
  J=4--3 and J=6--5 trace a different gas phase from that traced by
  the J=1--0, 2--1, 3--2 transitions,} i.e.  we find no region of the
$\rm (T_k, n, \Lambda _{co}) $ parameter space compatible with both
the observed values of $\rm r_{21}=(2-1){/}(1-0)$, 
$\rm r_{32}=(3-2){/}(1-0)$, $\rm R_{21}$, {\it
  and} the high $\rm r_{65/43}=(6-5)/(4-3)\sim 0.66\pm 0.26$ ratio
($\chi^2\ga 2.5$ over the entire parameter space).  The LVG solutions
derived solely from $\rm r_{21}$, $\rm r_{32}$ and $\rm R_{21}$
converge to $\rm T_k\sim (55-95)\, K$, $\rm n\sim 10^3\, cm^{-3}$
(best fit when $\rm [ ^{12}CO/ ^{13}CO]=100$, and for $\rm T_k=75\,K$;
$\chi^2=0.7$), and $\rm K_{vir}\sim 15\alpha ^{-1/2}$ (for $\rm
r_{co}= [CO/H_2]\sim 10^{-4}$).  This warm, non-virialised, gas phase
with $\rm \tau _{10}( ^{12}CO)\sim 1$ is typically found in ULIRGs, a
possible result of strongly evolving dynamical and far-UV-intense
environments (Aalto et al.  1995), but has expected ratios of $\rm
r_{43}\la 0.35$ and $\rm r_{65}\la 0.04 $, which are $\ga 2$ (for $\rm
r_{43}$) to $\ga 11$ (for $\rm r_{65}$) times lower than those
observed in Mrk\,231.  This result is illustrated in Fig.~3, where we
show the observed CO line luminosities (in solar units) together with
the values expected from the best-fitting LVG model derived from the
lowest three CO lines and the $^{13}$CO $J{=}2{-}1$ line only. The
fluxes of the high-$J$ lines far exceed the values predicted from this
model, demonstrating the presence of a distinct gas component, the
emission from which dominates the observed line fluxes.

\subsection{The dense gas phase}

The presence of a massive gas phase in Mrk\,231 that is much denser
than that dominating the low-J CO line emission can be inferred from
its luminous HCN J=1--0 line which makes this galaxy stand out even
amongst ULIRGs as the one with the highest HCN/CO J=1--0 luminosity
ratio ($\sim 0.25$, Solomon et al.  1992a).  The detection of HCN
J=4--3 ($\rm n_{crit}\sim 8.5\times 10^{6}\, cm^{-3}$) certainly
corroborates this, however a single line is not sufficient to point
uniquely to the presence of gas with $\rm n\ga n_{crit}$.  Much lower
densities are still possible if sub-thermal excitation and/or
radiative trapping (due to their considerable optical depths) were to
be significant.  Intensity ratios of widely-spaced HCN transitions are
excellent probes of the dense star-forming gas, and provide the key to
differentiating between different excitation processes, revealing a
considerable range of properties even in starbursts with similar FIR
and low-J CO luminosities (Jackson et al.  1995; Paglione, Jackson, \&
Ishizuki 1997).

The $\rm r_{43}(HCN)$ ratio measured in Mrk~231 indicates {\it a
  sub-thermally excited HCN J=4--3 line} which is well below that of
Arp\,220 ($\sim 0.8$; Greve et al.  2006), another archetypal ULIRG
often used as a typical template for high-z starbursts. The physical
conditions compatible with the dense-gas-dominated $\rm r_{43}(HCN)$
and HCN(1--0)/CO(6--5) ($=\rm R_{HCN/CO}$) brightness temperature
ratios were explored with our LVG code using HCN collisional rates for
the first 11 levels taken from the Leiden Atomic and Molecular
Database LAMDA.\footnote{http://www.strw.leidenuniv.nl/moldata/} The
value of $\rm R_{HCN/CO}$ helps discriminate over the considerable
range of conditions compatible solely with $\rm r_{43}(HCN)$, while
two additional constraints can be set by a) assuming that the HCN
emission emanates from virialized gas ``cells'' ($\rm K_{vir}\sim 1 $;
as is the case for the dense star-forming gas in the Galaxy), and b)
by stipulating that $\rm T_k\ga T_{dust}$.  The best solution ranges
found are those with $\rm T_k=40-45\,K$ and $\rm T_k=50-70\,K$ (see
Table 3), though the latter satisfies $\rm K_{vir}\sim 1$ only for a
$\sim 5-10$ times higher $[\rm HCN/H_2]$ abundance than that
considered typical for the Milky Way.  In starburst environments
values of $\rm T_k=50-70\,K$ may be possible because of a higher
ionization fraction of the molecular gas (Lepp \& Dalgarno 1996), or
higher C, C$^+$ abundances (which favor HCN production, Boger \&
Sternberg 2005) deeper inside molecular clouds caused by a more
vigorous turbulent diffusion of their atom-rich outer layers inwards.

The mass of the HCN-emitting gas phase can be estimated in a manner
similar to that used to determine the total molecular gas mass (using
the $ ^{12}$CO J=1--0 line), since the same arguments about line
emission from an ensemble of self-gravitating, non-shadowing (in space
or velocity), clouds apply.  Following Gao \& Solomon (2004a),

\begin{equation}
\rm M_{dense}(H_2)\approx 2.1 \frac{\sqrt{n(H_2)}}{T_{b}}\,
\left(\frac{M_{\odot}}{K\, km\, s^{-1}\, pc^2}\right)\, L_{HCN},
\end{equation}

\noindent
where  $\rm  T_b$ and  $\rm  L_{HCN}$  are the  area/velocity-averaged
brightness temperature  and the line luminosity of  an optically thick
HCN J=1--0 line (the case for  all solutions in Table~3).  For the two
sets  of conditions  that  best fit  the  dense gas  line ratios,  the
coefficient in  the equation above becomes  $\rm X_{HCN}\sim (19-20)\,
M_{\odot}\, (K\,km\,s^{-1}\,pc^{2})^{-1}  $ (for $\rm  T_k=40-45\,K $)
and      $\rm     X_{HCN}      \sim      (8-9)     \,      M_{\odot}\,
(K\,km\,s^{-1}\,pc^{2})^{-1} $ (for $\rm T_k=50-70\, K$).

The HCN J=1--0 luminosity of  Mrk\,231 is estimated using

\begin{equation}
\rm  L_{x}  =  \int  _{\Delta  V} \int_{A_s}  T_{b}\,da\,dV  =
\frac{c^2}{2 k_B \nu  ^2 _{x,rest}} \left(\frac{D^2 _L}{1+z}\right)
\int _{\Delta V} S_{\nu}\,dV,
\end{equation}

\noindent
where $\rm \Delta V$ and $\rm A_s$ are the total linewidth and the
area of the emitting source (where in this case x=HCN(1--0)) respectively.
Substituting and converting to astrophysically useful units yields

\begin{equation}
\rm L_{x} = 3.25\times 10^7\,(1+z)^{-1} 
\left(\frac{\nu _{x,rest}}{GHz}\right)^{-2} \left(\frac{D_L}{Mpc}\right)^2
 \left(\frac{\int _{\Delta V} S_{\nu}\,dV}{Jy\,km\,s^{-1}}\right)\, K\, km\,s^{-1}\,pc^2
\end{equation}

\noindent
For $\rm D_L$(z=0.042)=183.4\,Mpc and $\rm \nu _{x,rest}=88.63\,GHz$
(HCN J=1--0) the HCN J=1--0 velocity-integrated flux density (Table~2)
yields $\rm L_{HCN}=2\times 10^9\, K\, km\, s^{-1}\, pc^2$.  For the
range of $\rm X_{HCN}$ values derived previously, this corresponds to
$\rm M_{dense}(H_2) \sim (1.6-4)\times 10^{10}\, M_{\odot}$, while the
total H$_2$ gas mass estimated from the CO J=1--0 line luminosity of
$\rm L_{CO}=6.9\times 10^9\,K\,km\,s^{-1}\, pc^2$ ($\rm
\nu_{x,rest}=115.27\,GHz$, and the CO J=1--0 flux in Table~2), and a
standard Galactic conversion factor of $\rm X^{(Gal)}
_{CO}=4\,M_{\odot}\, (K\,km\,s^{-1})^{-1}$, is $\rm M_{tot}(H_2)\sim
3\times 10^{10}\, M_{\odot}$.  Thus, {\it at least $\sim 50\%$ and
  maybe all of the molecular gas mass in the ULIRG/QSO Mrk\,231 is
  dense ($\ga 10^4\,cm^{-3}$)}, quite unlike the state of the {\it
  bulk} of the molecular gas in our Galaxy and in low intensity
starbursts where $\rm \langle n(H_2)\rangle\sim 10^2-10^3\,cm^{-3}$
(e.g.  Paglione et al.  1997; Weiss, Walter, \& Scoville 2005).

The dynamical mass within a radius of $\rm R\sim 1.7\,kpc$, estimated
from high-resolution CO images, is $\rm M_{dyn}\sim 3.25\times
10^{10}\,M_{\odot}$ (Downes \& Solomon 1998, corrected for the adopted
cosmology).  Thus, either all that mass is molecular gas, or $\rm
M_{tot}(H_2)$ is overestimated by adopting an $\rm X^{(Gal)} _{CO}$
factor.  Extensive studies of ULIRGs suggest the latter because in
these extreme systems the molecular gas phase encompasses significant
amounts of non-gaseous mass (i.e.  stars) and thus one of the main
assumptions underlying a standard Galactic $\rm X_{CO}$ factor (that
of an ensemble of self-gravitating molecular clouds) breaks down.
These studies find $\rm X_{CO}\sim 1/5\, X^{(Gal)}_{CO}$ (Solomon
1997; Downes \& Solomon 1998), which for Mrk\,231 yields $\rm
M_{tot}(H_2)\sim 6\times 10^9\, M_{\odot}$, making it $\sim 2.5-6.5$
times smaller than $\rm M_{dense}(H_2)$ estimated from Equation~3.
This is clearly impossible, and argues for a revision also of the $\rm
X_{HCN}$ values derived from Equation 3.  Such a revision can be
understood in much the same terms underlying that of the $\rm
X^{(Gal)} _{CO}$ factor, since the dense gas phase seems
responsible for both HCN and most of the CO J=1--0 line emission. It
must be noted that this is not done for many HCN-deduced dense gas masses
in ULIRGs reported in the literature (e.g.  Solomon et al.  1992a) and
{\it thus these could be systematically overestimated by a factor
  of~$\sim $5.}

Interestingly, by adopting the same correction factor of $\sim 1/5$ for
$\rm X_{HCN}$ yields $\rm M_{dense}(H_2)\sim (3-8)\times 10^9\,
M_{\odot} $ for Mrk\,231 which, for its starburst-related IR
luminosity of $\rm L^{(*)} _{IR}\rm \sim 2/3\, L_{IR}$ (Downes \&
Solomon 1998) and $\rm L_{IR}(8-1000\mu m)=3.6\times 10^{12}\,
L_{\odot}$ (Sanders et al.  2003), raises the star formation
efficiency to $\rm \epsilon _{SF}=L^{(*)}_{IR}/M_{dense}(H_2)\sim
(300-800)\, L_{\odot}/M_{\odot}$, effectively bracketing the maximum
value of $\rm \sim 500\, L_{\odot}/M_{\odot}$ expected from O, B, star
radiation-feedback effects on the accreted dust/gas in star-forming
GMCs (Scoville 2004).  These may therefore be the hallmark features of
the star-forming molecular gas phase in galaxies -- {\it densities of
  $ n\ga 10^4\,cm^{-3}$, and star formation efficiencies of $\sim
  500\, L_{\odot}/M_{\odot}$} -- which in Mrk\,231 amounts to most of
its molecular gas mass.

\subsection{The diffuse gas phase}

Unlike the  HCN J=1--0,  4--3, and CO  J=6--5 line emission  which are
dominated by  the dense gas,  the lower-J CO lines  have contributions
also from a  diffuse phase.  The observed CO  (6--5)/(1--0) line ratio
can then be expressed as

\begin{equation}
\rm r_{65}=\frac{C_{ba}}{1+C_{ba}} r^{(b)} _{65},
\end{equation}

\noindent
where $\rm C_{ba}=f_{ba} T^{(b)} _{10}/T^{(a)} _{10} $ expresses the
contribution of emission from the dense phase (b) to that from a
diffuse and more extended phase (a), with $\rm f_{ba}<1$ being their
relative geometric filling factor, and $\rm T^{(a,b)} _{10}$ their CO
J=1--0 brightness temperatures (all quantities are velocity/area
averages).  All other CO line ratios where both phases contribute to
both transitions can be expressed as

\begin{equation}
\rm  r_{J+1\, J}=\frac{r^{(a)}  _{J+1\,  J} +  C_{ba} r^{(b)}  _{J+1\,
J}}{1+C_{ba}},\, \, with\, \, J+1=1,2,3,4.
\end{equation}

Typically we find $\rm C_{ba}\sim 0.6-0.7$ which, along with the known
range of $\rm r^{(b)} _{J+1\,J} $ values (obtained from the dense gas
properties outlined in Section 3.1), allows the subtraction of the
dense phase contribution from the observed ratios and the estimate of
the $\rm r^{(a)} _{J+1\,J} $ values.  These and the $\rm R_{21}\ga 37$
ratio (which we assume to be dominated by the diffuse phase) are then
used as inputs into our LVG code. This in turn yields $\rm n\sim
10^3\,cm^{-3}$ ($\rm T_{k}\sim 45-85\,K$) for the diffuse phase (a),
with still lower densities of $\rm n\sim 300\,cm^{-3}$ but $\rm
T_k\sim 80-140\,K$ also possible.  In all cases $\rm K_{vir}>1$
(reaching as high as $\sim 30$), which suggests the presence of highly
unbound gas whose large $\rm dV/dr$ values and high $\rm T_{k}$ are
responsible for its modest CO J=1--0 optical depths ($\rm \tau
_{10}\sim 0.5-1.5$).  This phase could be confined around individual GMCs
``enveloping'' their much denser self-gravitating regions, or could be
distributed over very different scales.  Interferometric CO line
imaging reveals a disk distribution with much of the lower-density
molecular gas in the outer regions (Downes \& Solomon 1998).  In
either case, the diffuse non self-gravitating gas phase is a minor
contributor to the total molecular gas mass of Mrk\,231.

\subsection{Thermal balance of the molecular gas in Mrk\,231}

The temperature of the molecular gas in a galaxy is determined by the
equilibrium between heating and cooling processes.  Gas heating is
dominated by the photoelectric effect on dust grains and polycyclic
aromatic hydrocarbons (PAHs), as discussed by e.g., Wolfire et
al.\ (1995).  Cooling proceeds through line radiation which, in normal
galaxies, is dominated by the [\cii] $158\mum$ line (e.g., Wolfire et
al.  1995, 2003; Kaufman et al.  1999), which can carry up to
$\sim0.1-0.5$\% of the total far-infrared (FIR) luminosity of a galaxy
(e.g., Stacey et al.\ 1991; Malhotra et al.\ 1997; Leech et al.\ 1999;
Pierini et al.\ 1999; Negishi et al.\ 2001).  In thermal equilibrium,
the heating and cooling rates are balanced: increased heating, such as
that resulting from enhanced star formation will be compensated by
increased cooling.  This has been used to estimate star formation
rates in nearby galaxies using the observed [\cii] line luminosity
(e.g., Crawford et al.\ 1985; Stacey et al.\ 1991; Boselli et
al.~2002).

Measurements obtained with the Long Wavelength Spectrograph (LWS) on
the Infrared Space Observatory (ISO) have shown that the [\cii]/FIR
flux ratio decreases in galaxies of very high FIR luminosity, i.e.,
the increase in [\cii] line luminosity is no longer proportional to
the FIR luminosity. This effect was first discovered in deep ISO
measurements of a small sample of nearby ULIRGs (Luhman et al.\ 1998),
where the observed [\cii]/FIR luminosity ratio was found to be smaller
than $\sim0.05$\% in some of the most FIR-luminous galaxies - a value
more than factor of 10 lower than that observed in less luminous
galaxies.  Various explanations for this effect have been proposed,
including dust absorption and saturation effects, self-absorption in
the [\cii] line and more subtle explanations related to the detailed
physics of dense photon dominated regions (PDRs) (Luhman et
al.\ 2003).

The high luminosity  of the [\cii] line suggests  that the line should
be observable  (redshifted into the submillimeter regime)  out to very
high redshifts (e.g., Loeb 1993; Stark 1997; Suginohara et al.\ 1999).
Indeed, ALMA  will be able to  detect the [\cii] emission  from a Milky
Way-type galaxy out to $z\sim5$  (Van der Werf \& Israel 1996).  It is
therefore imperative to understand the physical conditions determining
the strength  of the  [\cii] line more  fully.  The first  attempts to
detect this line in the distant  Universe were carried out by Isaak et
al.\ (1994)  on the $z=4.7$  QSO BR\,1202$-$0725, with an  upper limit
implying a [\cii]/FIR ratio lower  than in the nearby starburst galaxy
M82.  In a  deeper integration on this same object,  an upper limit on
the [\cii] luminosity was found which is less than 0.06\% of the total
FIR luminosity (Van der Werf 1999).  Recently, the first detections of
[\cii]  at  high redshift  have  been  obtained  in the  $z=6.42$  QSO
SDSS\,J1148+5251  (Maiolino et  al.\  2005), and  in  the $z=4.7$  QSO
BR\,1202$-$0725  (Iono et al.,  2006) where  the [\cii]  luminosity is
$0.02-0.04$\% of the total far-infrared luminosity.  All these results
are consistent with ISO observations  of local ULIRGs, and suggest that the
same physical mechanisms are responsible.

Our present results for Mrk\,231 amount to the most complete picture
of the molecular gas attained for a local ULIRG enabling us to analyze
the cooling budget of its molecular medium in detail, and thereby shed
light on the [\cii] problem. The results are summarized in Table~4
where we have used the mean values from our LVG models to calculate
luminosities for CO lines which have not been observed. The HCN lines
(not relevant for the cooling budget) resulting from the same models
are presented in Table~5. The [\cii] $158\,\mu$m line luminosity has
been derived from the line flux of $3.2\pm0.4\cdot
10^{-20}\,$W\,cm$^{-2}$ measured using the ISO LWS by Luhman et
al.\ (1998).  The cooling due to the $370\,\mum$ and $609\mum$ [\ci]
lines was estimated from a measurement of the latter (Gerin \&
Phillips 2000) and by assuming LTE (the [\ci] energy levels are
expected to be fully thermalized for the dense gas phase).  The
$T_{\rm k}$ range of the best two LVG solution ranges (Table~3)
constrains the CI(2--1)/(1--0) brightness temperature ratio to $\sim
0.8-1.2$, for which we adopt the mean value of $\sim 1$ (which is
actually measured in M\,82; Stutzki et al.  1997).  We also list the
total luminosity $L_{\rm dense}$ in CO lines up to $J{=}10{-}9$ for
the dense phase only, as well as those the total luminosity $L_{\rm
  diffuse}$ in CO lines from the diffuse phase, calculated from the
difference between observed fluxes and the modeled fluxes from the
dense phase.  Line strengths are presented in flux units ($S_{\rm
  line}$ in Jy\,km\,s$^{-1}$), in luminosity units ($L_{\rm line}$ in
$L_{\odot}$), and in $L'_{\rm line}$ luminosities (in
K\,km\,s$^{-1}$\,pc$^2$, estimated using Eq.~5); the last quantity
scales directly with the intrinsic brightness temperature of the line,
and is thus constant for thermalized optically thick lines originating
from the same medium.

Table~4 shows  the remarkable result  that the cooling  luminosity for
the  dense molecular gas  in Mrk\,231  considering only  CO approaches
that of the [\cii] line, with $L_{\rm CO}/L_{\rm FIR}=1.1\pm0.2\cdot10^{-4}$
while  $L_{[{\rm  C\,II}]}/L_{\rm  FIR}=1.5\pm0.2\cdot10^{-4}$, for  $L_{\rm
FIR}(40-400\,\mu{\rm m})\sim 2.3\cdot10^{12}\,L_{\odot}$  in   Mrk\,231
(Sanders et al., 2003 for  the cosmology adopted here).  The situation
is further illustrated in Fig.~4, where we show a bar histogram of the
cooling lines of the interstellar  gas in Mrk\,231. Error bars are
observational errors for the luminosities based on measured fluxes, and
indicate the range allowed by the LVG models for the remaining values.
For the latter values we adopted a minimum uncertainty of at least 30\% based
on the measurement errors of the CO $J{=}6{-}5$ and $4{-}3$ lines. The
error bars thus give a good indication of the expected ranges taking
into account uncertainties in both the observations and the LVG modeling.
It is instructive
to compare  these results to  the corresponding results for  the Milky
Way, where  global CO line fluxes have been measured  using the
COBE  data, and converted into luminosities using a model for the
spatial distribution of the emission within the solar circle by
Wright  et al.\ (1991).  For  the Milky  Way the  total CO
cooling   (relative   to  FIR   luminosity)   is  $L_{\rm   CO}/L_{\rm
FIR}\sim2\cdot10^{-5}$,  while  for  [\cii] $L_{[{\rm  C\,II}]}/L_{\rm
FIR}=2.8\cdot10^{-3}$,  i.e.,  the  cooling  is totally  dominated  by
[\cii]  with neglible  CO  cooling.   The importance  of  the CO  line
cooling is also reflected in  the CO line ratios: the flux ratio
CO $J{=}4{-}3{/}J{=}2{-}1$  (which has been observed  both objects) is
1.5 in the Milky Way (Wright  et al., 1991), but 5.4 in Mrk\,231. We
can also compare these results to  a recent survey in mid-$J$ CO lines
and the two [\ci]  lines by Bayet et al.\ (2006) of  a small sample of
local starburst galaxies.   In these objects the mid-$J$  CO lines are
also found to be much stronger  than in the Milky Way, with cooling by
CO lines significantly exceeding that  from [\ci]. However, CO cooling in these
objects  is  still  insignificant  compared  to  [\cii], in
contrast to the much more extreme results obtained here for Mrk\,231.

The cooling  associated with  the diffuse molecular  gas (contributing
only to  the lowest  three CO transitions)  of Mrk\,231  is relatively
unimportant.  Thus it is natural  to identify the dense phase with the
actively star  forming gas,  which experiences the  strongest heating,
and must  therefore also cool  efficiently.  Unlike the Milky  Way and
lower intensity starbursts in Mrk\,231 this phase contains most of the
molecular gas mass,  forming stars at what is thought  to be a maximum
efficiency (Section  3.1), at rates  found only in  dense star-forming
cores in the Milky Way.

The  resulting picture  of a  dense and  dominant (in  terms  of mass)
molecular gas  phase, emitting strongly  in CO $J{=}6{-}5$ but  with a
suppressed [\cii]  line, points towards  dense PDRs (e.g.,  Kaufman et
al., 1999). In  such PDRs, the high density leads  to a high formation
rate for  CO (proportional to  $n^2$), while the  CO photodissociation
rate is  less strongly enhanced  (proportional to $n$).  As  a result,
the  ionized carbon  layer is  thin,  leading to  a suppressed  [\cii]
line. In addition, the transition from ionized and atomic carbon to CO
now takes place  closer to the source  of heating, resulting
in a large  column density of warm molecular  gas with strong emission
in the mid-$J$  CO lines. Generalizing to ULIRGs  as a population, our
Mrk\,231  result  suggests that  {\it  the  suppressed\/} [\cii]  {\it
cooling in  ULIRGs is  a result  of high densities  for most  of their
molecular gas,  bathed in strong  far-UV radiation fields,}  quite unlike
lower intensity starburst or quiescent galaxies where much more modest
amounts of molecular gas reside in such a high density phase.

Our  explanation   predicts  strong  emission   in  CO  lines   up  to
$J{=}10{-}9$  in  ULIRGs.  While   such  high-J  transitions  are  not
observable from the  ground, they will be accesible  with HIFI onboard
the Herschel satellite.  Observations  of high excitation CO lines from
space or the  ground, as well as careful estimates  of the fraction of
molecular  gas mass  with $n\geq 10^4\,$cm$^{-3}$  in  local ULIRGs
using HCN transitions, will provide  a critical test of the hypothesis
that dense  PDRs lie  at the  heart of the  [\cii] cooling  problem in
ULIRGs.

\section{Molecular lines at high redshifts: caveats and expectations}

The wealth of molecular line and dust continuum data used to constrain
the state of molecular gas in Mrk\,231 and assign most of its mass to
a dense phase is rarely available for objects in the distant Universe.
Typically, two high-$J$ CO lines are detected (Solomon et al.~1992b;
Solomon \& Vanden Bout 2005 and references therein), with a similar
sparsity of dust continuum measurements.  The few frequencies observed
(usually at the Rayleigh-Jeans part of a SED) provide poor estimates
of $T_{\rm dust}$, which in turn provides little constraint on the
range of $T_{\rm k}$.  In most cases, the often (but not always) large
CO line ratios of high-$J$ lines measured in high redshift objects are
interpreted in terms of a warm phase, yet infact they trace a wider
range of physical conditions.  This degeneracy can be easily
demonstrated by setting the high $\rm r_{65/43}$ ratio measured in
Mrk\,231 as the only constraint available.  It can then be seen
(Table~6) that besides the typical warm gas phase with moderate
optical depths, gas as cold as $\rm T_k=15\,K$ but $\sim 30-100$ times
more dense also reproduces the high $\rm r_{65/43}$ ratio that we have
observed. In such a phase the large CO line optical depths, and the
resulting radiative trapping, thermalize transitions up to high $J$
levels so that
 
\begin{equation}
\rm r_{65/43}\sim \frac{T_{65}}{T_{43}}\left( \frac{e^{T_{43}/T_k}-1}{e^{T_{65}/T_k}-1}\right)
=\frac{3}{2} \left( \frac{e^{22/T_k}-1}{e^{33/T_k}-1}\right),
\end{equation}

\noindent
($\rm T_{J+1\, J}=h\nu _{J+1\,  J}/k_B$), which for e.g. $\rm T_k=17\,
K$ yields $\rm r_{65/43}\sim 0.66 $ (the observed value).

The degeneracies alluded to above reflect real conditions found in
GMCs: warm, dense gas associated with star formation, and
dense-but-cold gas in regions with no such significant activity.  In
ULIRGs a massive, dense but cold phase is unlikely but not impossible
during the rapid dynamic evolution expected for mergers (e.g., Aalto
2005).  Observing at least one $ ^{13}$CO transition is pivotal to
discriminating between cold dense gas with an optically thick CO
J=1--0 line and a less dense warmer phase where $\rm \tau _{10}(
^{12}CO)\la 1$.  Indeed, the cold/dense solutions in Table 6
correspond to $\rm ^{12}CO/ ^{13}CO$ J=1--0 intensity ratios of $\rm
R_{10}\sim 1-2$, much lower than those even in the coldest GMCs ($\sim
5$).  Interestingly, setting $\rm K_{vir}\ga 1$ as a constraint
selects the conditions with $\rm T_k\ga 40\,K$ (Table 6), and thus may
also be useful in ``breaking'' such degeneracies, though it will
always depend on the assumed abundances.

\subsection{Molecular gas SLEDs for Mrk\,231: 
towards establishing local benchmarks}

High resolution imaging of molecular lines and their relative
strengths with the next generation of mm/sub-mm arrays holds the key
to unobscured views of deeply dust-enshrouded star forming regions in
galaxies, their dynamical masses, and the molecular gas fueling the
embedded star formation.

This is apparent now that CO-bright H$_2$
gas has been detected out to $\rm z\sim 6$ (Walter et al.  2003), and
distant dust-enshrouded optically faint starbursts, responsible for
building large fractions of the stellar mass at present epochs, have
been discovered (e.g.  Smail, Ivison, \& Blain 1997).  
The need for
local molecular line SLEDs for the bulk of the molecular gas in
galaxies (and not just for their starburst sub-regions) is underlined
by the fact that currently there are more detections of CO J=4--3 or
higher-J lines in high redshift rather than local starbursts (e.g.
Solomon \& Vanden Bout~2005).

The compact distribution of the CO, HCN emission in local ULIRGs and
their large IR and molecular line luminosities makes them ideal
objects for establishing such SLEDs -- single pointings encompass all
the H$_2$ distribution -- a situation mirrored in high-z galaxies.
The extreme-starburst/QSO present in Mrk\,231 makes this galaxy in
particular an often- used benchmark for similar high-z systems
(e.g. Wagg et al.  2005).  We use the constraints on its dense gas
properties to derive the expected strengths of the unobserved CO and
HCN transitions.  The resulting CO and HCN SLEDs can then be used to
assess the capabilities of the next generation cm (EVLA, SKA) and
mm/sub-mm (ALMA) arrays, as well as the spaceborne Herschel
Observatory in detecting star-forming molecular gas throughout the
Universe (e.g. Van der Werf \& Israel 1996). 
The line fluxes, calculated for a range of redshifts, are
presented in Fig.~5, where we also show the analogous values for the
Milky Way, with luminosites again from the COBE data (Wright et al.,
1991) for all lines except CO $J{=}1{-}0$. The luminosity of the
latter line in the Milky Way was estimated using the CO $J{=}2{-}1$
luminosity from Wright et al.\ (1991) and the CO $2{-}1/1{-}0$ line
ratio in the inner $2.5^\circ$ of the Milky Way as measured by COBE,
which is within the errors equal to the global line ratio over the
inner Galaxy (Fixsen, Bennett, \& Mather, 1999).  The Milky Way and
Mrk\,231 bracket the two extremes of quiescent and starburst galaxies.
A luminosity of $L_{\rm FIR}=2\cdot10^{10}\,L_{\odot}$ (Wright et al.,
1991) makes the Milky Way detectable with ALMA out $z\sim1$, using CO
$J{=}3{-}2$ in the $211-275\,$GHz band.  Higher CO lines will not be
detectable however, though redshifted [\cii] will be observable with ALMA
out to $z\sim4-8$ with significant integration time.  In the case of
a Mrk\,231-like object, the situation is quite different, with
transitions up to CO $J{=}9{-}8$ detectable with ALMA out to
$z\sim10$.  The two [\ci] lines can be detected out to $z\sim2.5$, and
the [\cii] line will be readily detectable at $z\sim4-8$.  The
high-density tracing HCN lines will typically be detectable only out
to $z\sim1.5$, where the $84-116\,$GHz band of ALMA will be the most
suitable frequency range.

It is instructive to investigate how existing detections of CO lines at high
redshift fit on the Mrk\,231 template.  We use the comprehensive
compilation by Solomon \& Vanden Bout (2005; their Appendix~2) of all
detections of (sub)millimeter lines from high-$z$ galaxies to 2005,
supplemented with more recent data by Riechers et al.\ (2006).  In
Fig.~6 we plot again the CO cooling budget of Mrk\,231 but now overlay
the observed line luminosities of all high-$z$ objects with at least
two detected CO lines (one of them being CO $J{=}3{-}2$ or
$J{=}2{-}1$). Ideally all fluxes should be scaled relative to one
common line, but unfortunately there is no line which is detected in
all (or even most) objects. However, almost all high-$z$ objects with
CO detections have either the $J{=}3{-}2$ or the $J{=}2{-}1$ line
detected (but, remarkably, so far never both). We therefore scale the
observed line fluxes to either the CO $J{=}3{-}2$ or the CO
$J{=}2{-}1$ luminosity of Mrk\,231 and show the results in
Fig.~6. Inspection of this figure reveals a considerable range in
$J{=}4{-}3{/}2{-}1$ line ratios, and it is clear that interpolation to
a fictitious $J{=}3{-}2$ value would have introduced significant
uncertainties. Therefore we prefer to normalize to either the $J{=}3{-}2$
or the $J{=}2{-}1$ line, which, although blurring the comparison
somewhat, has the advantage of being based on measurements rather than
an uncertain interpolation. Physically, this
procedure is acceptable since these lines have similar contributions
from the dense gas component (70\% for the $2{-}1$ line, 80\% for the
$3{-}2$ line) in our Mrk\,231 fiducial model.

Figure 6 shows that Mrk\,231 is a reasonable template for most
high-$z$ objects that have been detected in CO, with line ratios
within a factor of two from those of Mrk\,231 up to about $J{=}6{-}5$.
This does not imply that {\it all\/} high-$z$ objects would have
similar line ratios: there is clearly a selection effect in the sense
that high-$z$ objects are followed up with low-$J$ CO observations
only after having being detected via their luminous high-$J$ line
emission.  With this caveat in mind, Mrk\,231 appears to be a useful
template for high-$z$ galaxies detected in CO to-date. It is
interesting to note that this is the case also for its dust continuum
emission (see e.g., Fig.~2 of Blain et al., 2002).

Two notable outliers in Fig.~6 are HR10 and APM\,08279+5255. HR10 is
an extremely red object at a modest redshift of $z=1.439$ which, while
a ULIRG ($L_{\rm FIR}\sim6\cdot10^{12}\,L_{\odot}$), has line ratios
indicative of low excitation conditions more similar to the Milky Way
than to Mrk\,231 (Papadopoulos \& Ivison, 2002).  {\it Detections and
  imaging of such low excitation objects beyond $z\sim4$ (even with
  such large $L_{\rm FIR}$) will be difficult even with ALMA\null}.
At the other extreme, APM\,08279+5255 shows unusually high excitation,
with CO line luminosities rising all the way up to at least
$J{=}10{-}9$: gas at high gas densities is indicated by the detection
of luminous HCN $J{=}5{-}4$ by Wagg et al.\ (2005). While these
authors can fit their data with a single
gas phase of $n({\rm H}_2)\sim 4\times
10^4\,$cm$^{-3}$, this object is strongly lensed (Lewis et al., 2002),
and so it is likely that the observed fluxes are dominated by a
selectively amplified region of high-density molecular gas. Another
indication of a singularly different molecular line excitation of this
distant starburst/QSO is its HCN(5--4)/CO(4--3) luminosity ratio of
$L'_{\rm line}({\rm HCN})/L'_{\rm line}({\rm CO})\sim 0.21-0.36$ (from
Wagg et al.\ 2005 and Downes et al.\ 1999), while from the observed CO
J=4--3 and the expected HCN J=5--4 luminosity (deduced from the best
LVG solutions for the dense gas) the much lower values of
$\sim 0.045-0.066$ are derived for Mrk\,231.

Agreement with (or deviation from) the Mrk\,231 SLED appears to depend
primarily on the mean gas density of the phase dominating the
emission.  This is suggested by further inspection of Fig.~6 in
relation to the HCN line luminosities where available.  In
IRAS\,F10214+4724, VCV\,J140955.5+562827, and the Cloverleaf quasar,
HCN J=1--0 has been detected, but with an HCN(J=1--0)/CO(J=3--2) ratio much
lower than in Mrk\,231, and these galaxies reveal somewhat lower CO
line excitation than Mrk\,231 as well.  In contrast, the
HCN(J=2--1)/CO(J=2--1) ratio in SDSS\,J1148+5251 is higher than in
Mrk\,231, and this object shows also higher excitation in the CO
lines.  Construction of molecular gas SLEDs out to the high
excitation regimes associated with the star-forming gas for a large
number of local LIRGs will allow a more thorough examination of such
differences, while establishing much-needed benchmarks for the
interpretation of the still sparse data at high redshifts.

\section{Conclusions}

We report  on the detections of  the CO J=6--5,  J=4--3 transitions of
the archetypal  nearby ULIRG/QSO Mrk\,231.  These  first detections of
high-excitation  CO lines  in a  local extreme  starburst,  along with
hereby reported CO J=3--2, HCN  J=4--3 detections and CO J=1--0, 2--1,
$  ^{13}$CO   J=2--1,  HCN   J=1--0  literature  data,   allow  robust
constraints  to  be  placed  on  its molecular  gas  properties.   Our
conclusions can be summarized as follows:

\begin{enumerate}
\item
The high excitation CO J=4--3 and J=6--5 lines trace denser ($\rm \ga
10^4\,cm^{-3}$), gas phase different from that dominating the lower
three CO transitions.  This phase is also responsible for the observed
HCN line emission, and contains $\ga 50\%$ of the total molecular gas
mass in this galaxy, quite unlike quiescent or lower intensity
starbursts where only a few percent of the total molecular gas mass
reside in such a phase.  This dominant dense gas phase fuels the
starburst in Mrk\,231 at almost the maximum expected star forming
efficiency of $\rm L^{(*)} _{IR}/M(n\ga 10^4\,cm^{-3})\sim 500\,
L_{\odot }/M_{\odot}$.
\item
This preeminence of dense  and relatively warm ($\rm T_k\sim 40-70\,
K$) gas in Mrk\,231 presents  a totally different thermal balance from
that found  in more quiescent  galaxies, with CO comparable  to [\cii]
emission line  cooling.  If  confirmed, through observations  of other
ULIRGs, this may provide the explanation of the long-standing issue of
their   very   low   [\cii]/IR   luminosity  ratios   namely,   bright
high-excitation CO lines from dense far-UV photon-dominated molecular gas
are the major  and even dominant coolants with  respect to their faint
[\cii] line.
\item
Since the excitation conditions probed
by the mid-$J$ CO lines and 
low-excitation atomic and ionic fine-structure lines are very
similar, mid-$J$ CO lines provide an excellent 
diagnostic complementary to e.g., PACS on the ESA Herschel satellite, 
and proposed future missions such as the Japanese SPICA
project. With ALMA these diagnostics will be available at
unprecedented spatial resolution.
\item
The well-constrained  dense molecular gas SLED of  Mrk\,231 provides a
very  valuable  template  for  direct comparisons  with  starburst/QSO
systems  at  high  redshifts, with  the  caveat  that  it may  not  be
representative  of the ULIRG  population as  a whole  (e.g.  Mrk\,231,
unlike  Arp\,220,  has  a  sub-thermal HCN(4--3)/(1--0)  line  ratio).
Nevertheless it  is a template  unaffected by the  strong differential
lensing  that may  be  skewing  the intrinsic  SLEDs  of several  high
redshift  galaxies towards  those of  a more  compact, highly-excited,
star-forming and even AGN-related molecular gas phase.
\end{enumerate}

\acknowledgments We  should like  to thank the  superb crew  of people
supporting the operation of the James Clerk Maxwell Telescope. Special
thanks to  Iain Coulson, Jim Hoge,  and Per Friberg  for assisting and
advising  us on  a  demanding  set of  observations.  PPP thanks  Axel
Wei\ss\,  for  helpful  comments   and  suggestions  on  the  original
manuscript.   KGI would  like to  remember George  Isaak, for  all the
support,  encouragement, inspiration and  very heated  discussion over
the years that only a father can give.\\

Facilities: \facility{JCMT}\\

\clearpage

\begin{figure}
\plotone{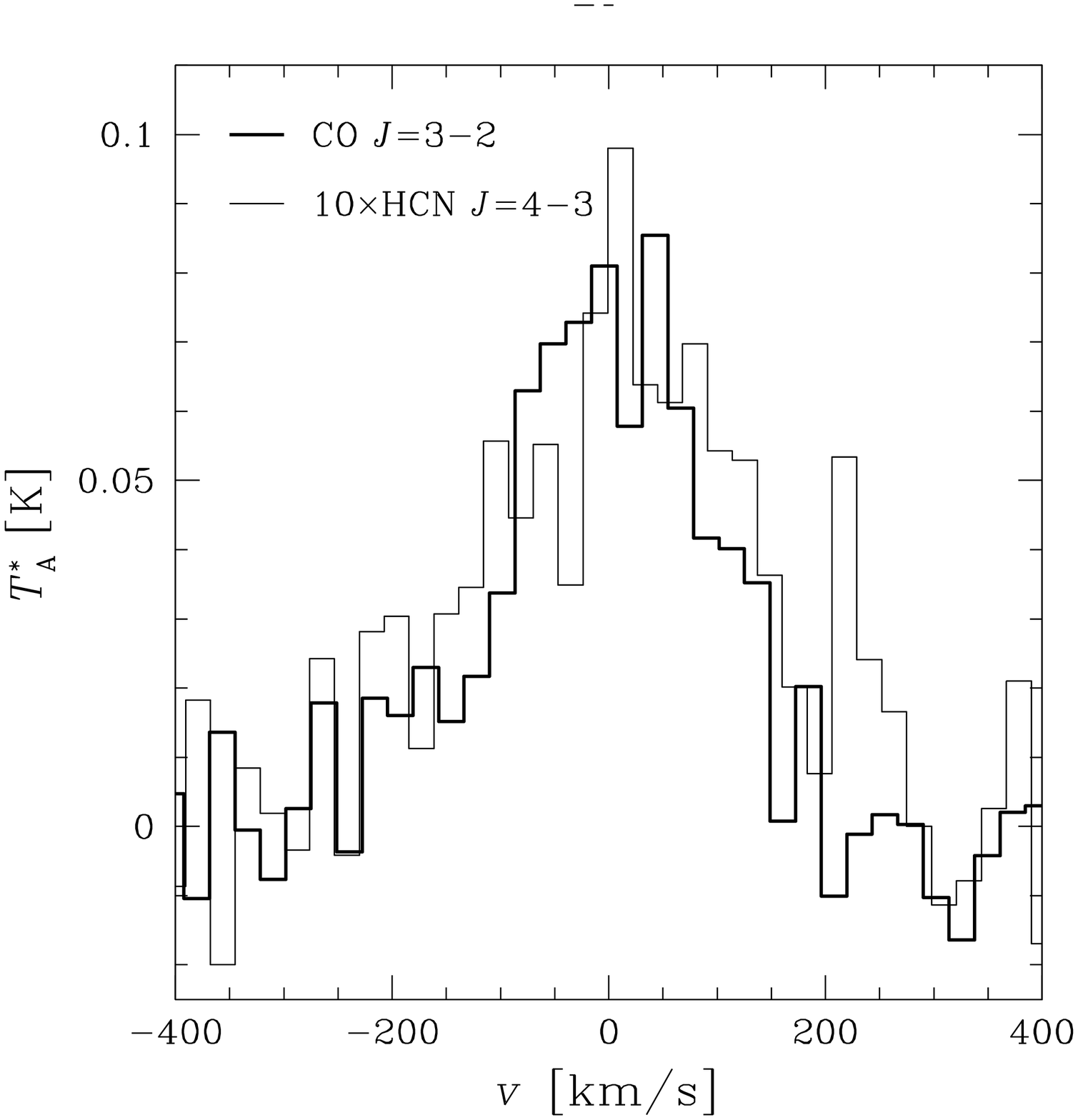}
\caption{Mrk\,231: $\rm \alpha  = 12^h\  56^{m}\  14.18^s$,  $\rm
\delta = +56^{\circ}\  52^{'}\ 25.8^{''}$ (J2000). 
The  CO J=3--2
(thick line), 10$\times  $[HCN J=4--3]  (thin line)  spectra  at 
 resolution $\rm \Delta \nu _{ch}=25\,
MHz$ ($\rm  \sim 21.6\, km\, s^{-1}$),  with thermal rms errors of
$\rm \delta T^* _A\sim 9\, mK$ (CO J=3--2), and $\rm \delta T^* _A\sim
1\, mK$ (HCN J=4--3).
Velocities are relative 
to $v_{\rm LSR}=12650\,$km\,s$^{-1}$.}
\end{figure}

\clearpage

\begin{figure}
\plotone{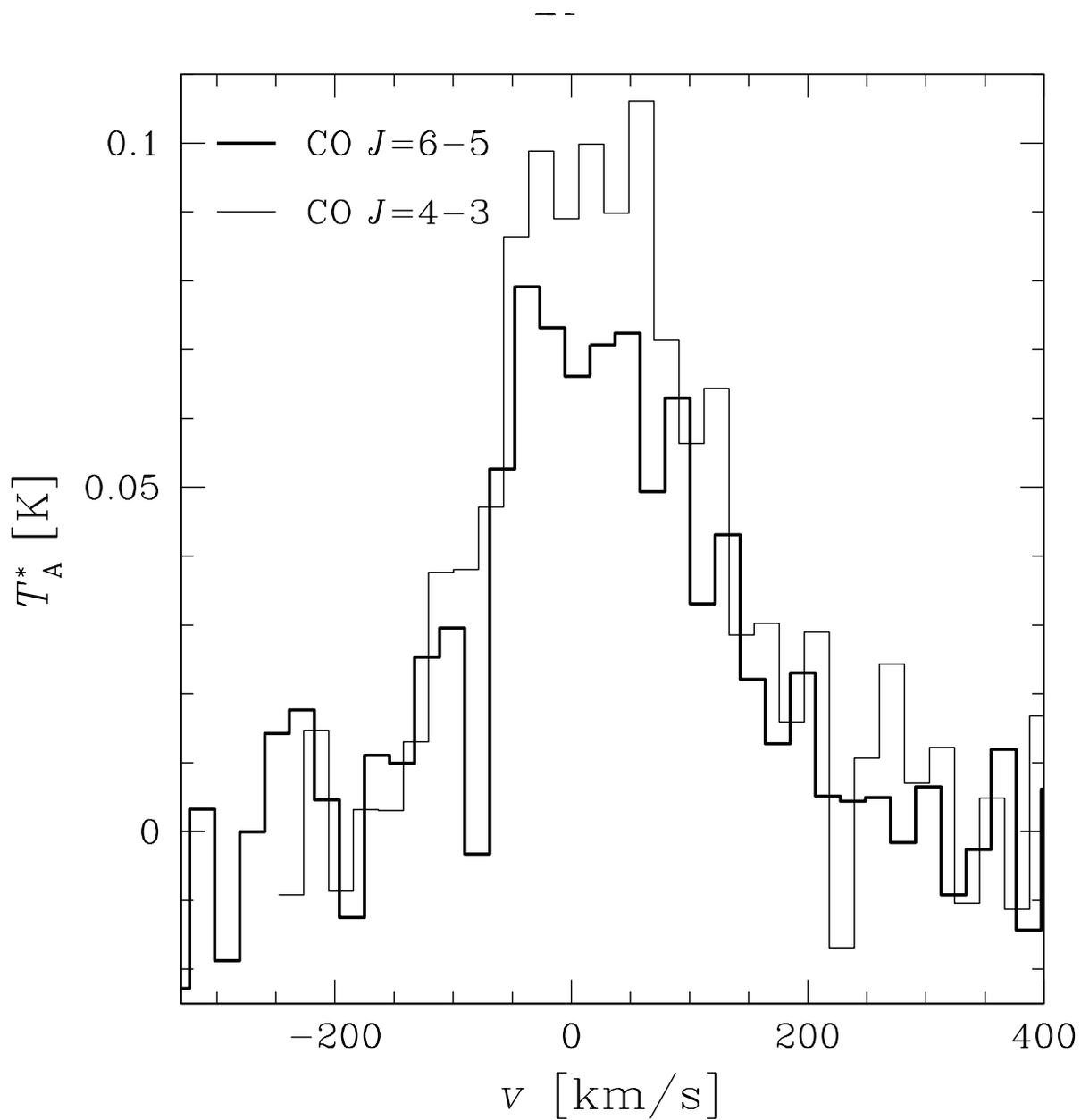}
\caption{High excitation CO transitions: CO J=6--5 (thick line) and CO
J=4--3  (thin line), at  a common  resolution of  $\rm \sim  20\, km\,
s^{-1}$.  Thermal  rms errors: $\rm  \delta T^* _A(4-3)\sim  12\, mK$,
and $\rm \delta T^* _A(6-5)\sim 15\, mK$.
Velocities are relative 
to $v_{\rm LSR}=12650\,$km\,s$^{-1}$.}
\end{figure}

\clearpage

\begin{figure}
\epsscale{1.0}
\plotone{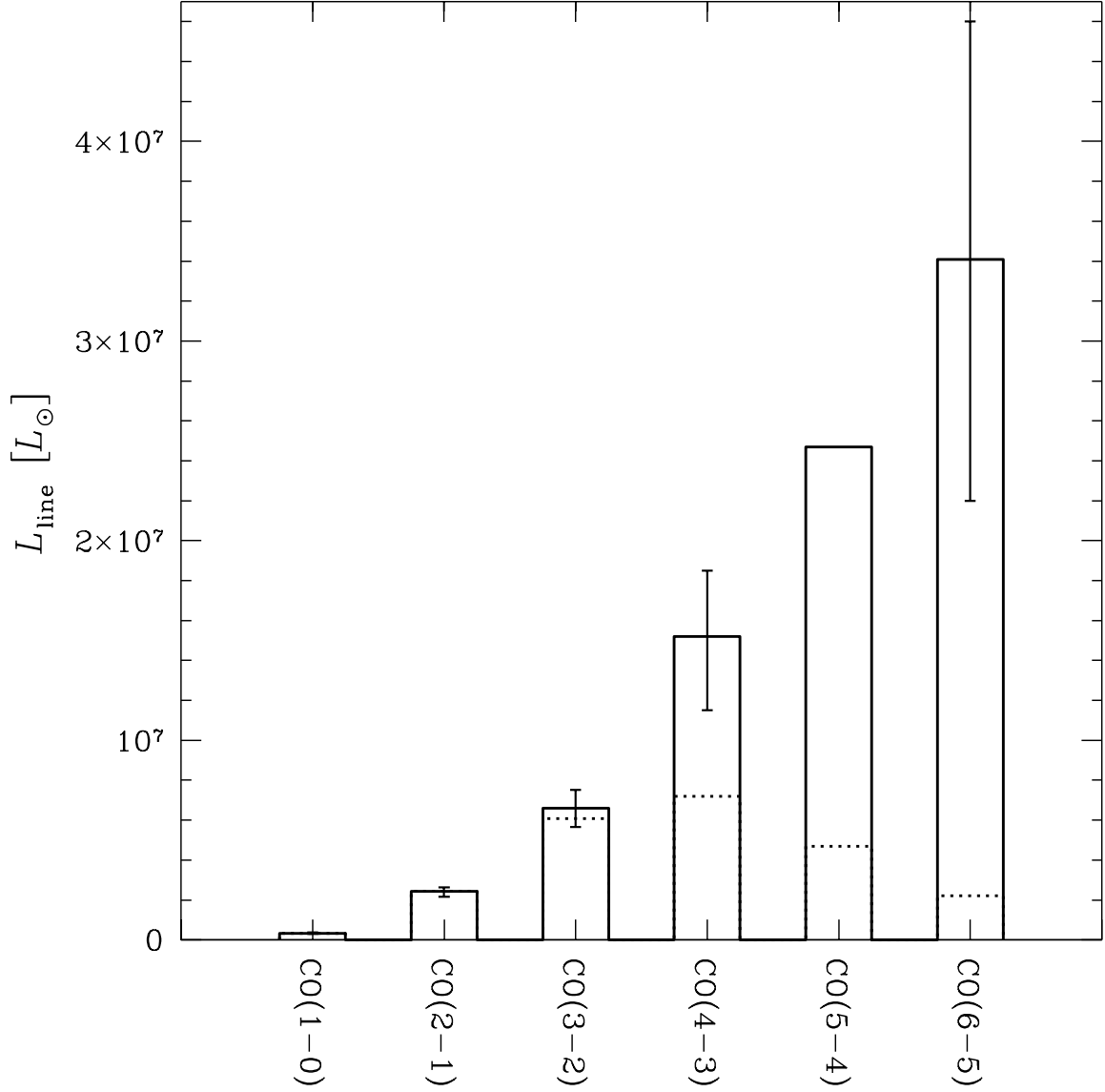}
\caption{Observed luminosities of CO $J{=}1{-}0$, $2{-}1$, $3{-}2$, $4{-}3$
  and $6{-}5$ (and interpolated value for CO $J{=}5{-}4$) shown by the drawn
  boxes with error bars; the dashed lines give the expected luminosities
  based on the best-fitting single component  LVG model derived 
  considering only the three lowest
  CO lines and $^{13}$CO $J{=}2{-}1$.}
\end{figure}

\clearpage

\begin{figure}
\epsscale{1.0}
\plotone{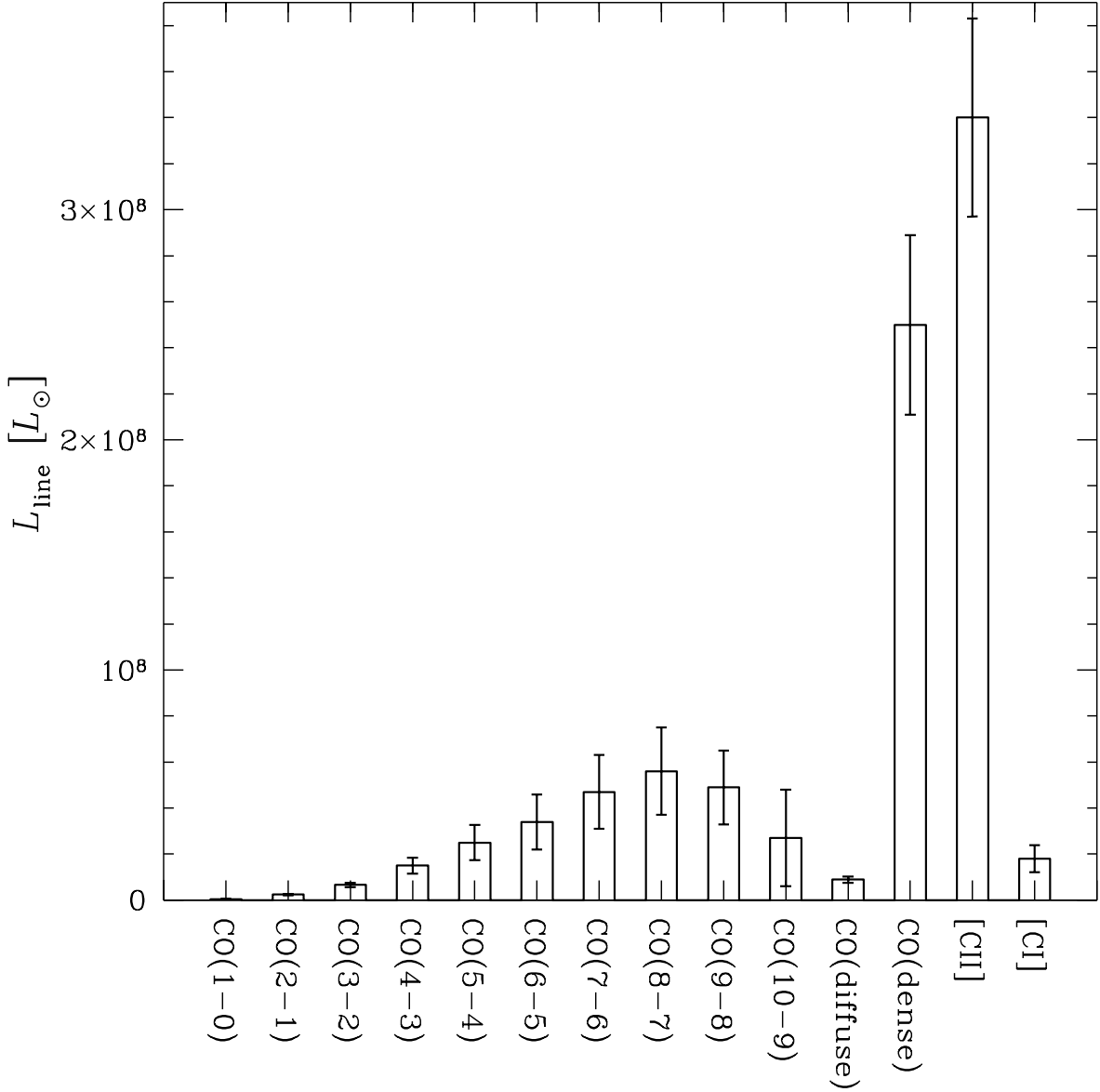}
\caption{Cooling  budget  of  the  molecular  gas  of  Mrk\,231.
  Plotted are the total CO line and [\cii] line luminosities 
  from  Table~4, which represent the sum of the diffuse and dense gas
  phases. The bar for [\ci] represents the
  sum of the expected luminosities of the two [\ci] lines (see text).
Error bars are
observational errors for the luminosities based on measured fluxes, and
indicate the range allowed by the LVG models for the remaining values.
For the CO lines deduced from these models
we use a minimum uncertainty of 30\% (based
on the measurement errors of the CO $J{=}6{-}5$ and $4{-}3$ lines).
}
\end{figure}

\clearpage

\begin{figure}
\plotone{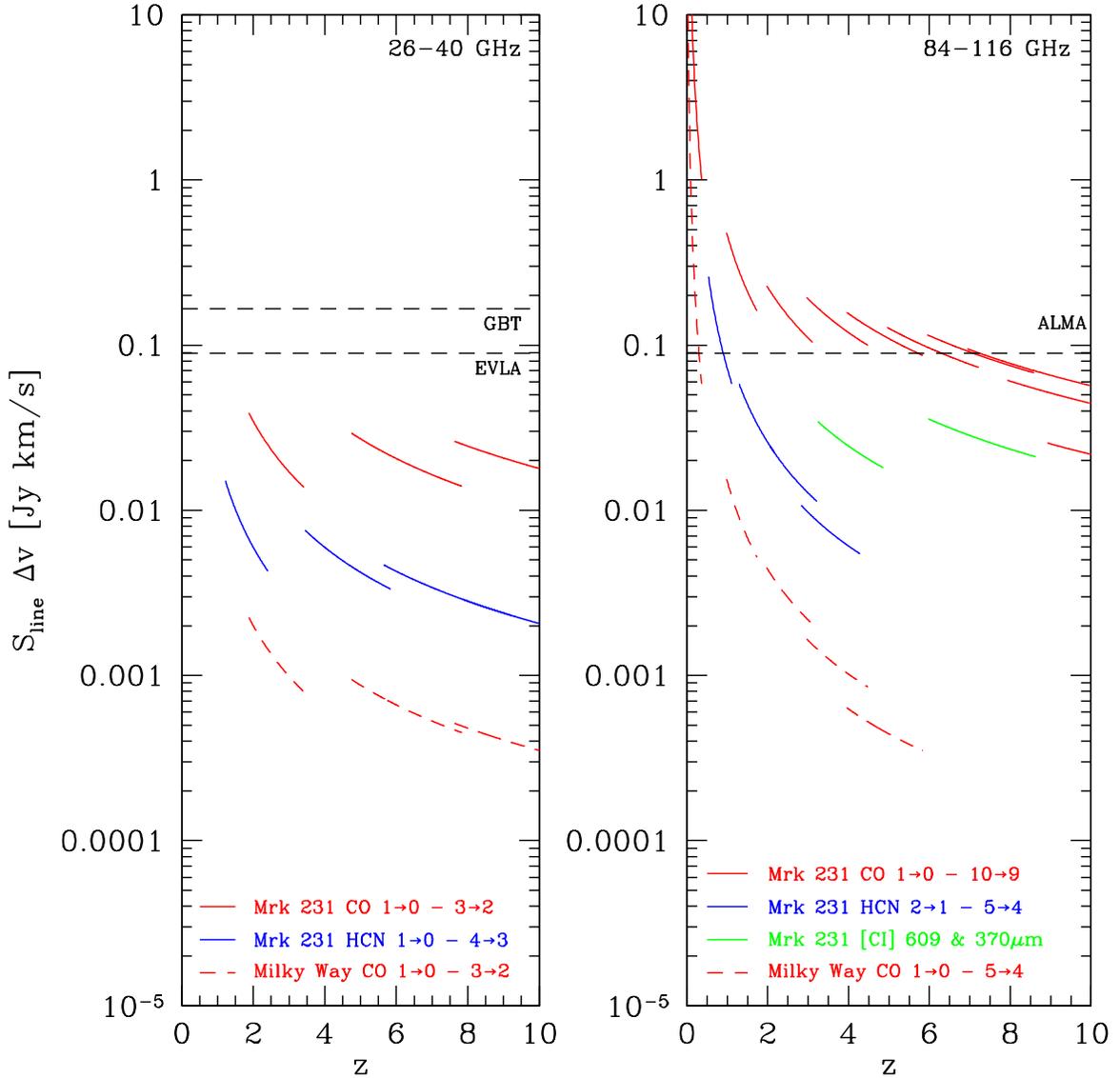}
\caption{Predicted fluxes of Mrk\,231 and  the Milky Way as a function
of redshift.   The different panels  show the lines redshifted  into 4
different  frequency  intervals,  as  indicated  at the  top  of  each
panel. Also indicated are instrumental sensitivity limits (5\,$\sigma$
point   source  limits  in   1~hour  at   a  velocity   resolution  of
300\,km\,s$^{-1}$; it is assumed that  the full line flux is contained
in this  velocity interval). These  limits have been  calculated using
the      on-line       sensitivity      calculators      for      ALMA
(http://www.eso.org/projects/alma/science/bin/sensitivity.html) using 50
antennas, and
GBT    (http://www.gb.nrao.edu/GBT/setups/senscalc.html),    and   are
projected  values for  the  EVLA\null.  These  figures  assume a  flat
$\Lambda$-dominated  cosmology with $H_0=71\,$km\,s$^{-1}$\,Mpc$^{-1}$
and  $\Omega_{\rm  m}=0.27$.}
\end{figure}

\clearpage

\begin{figure}
\figurenum{5}
\plotone{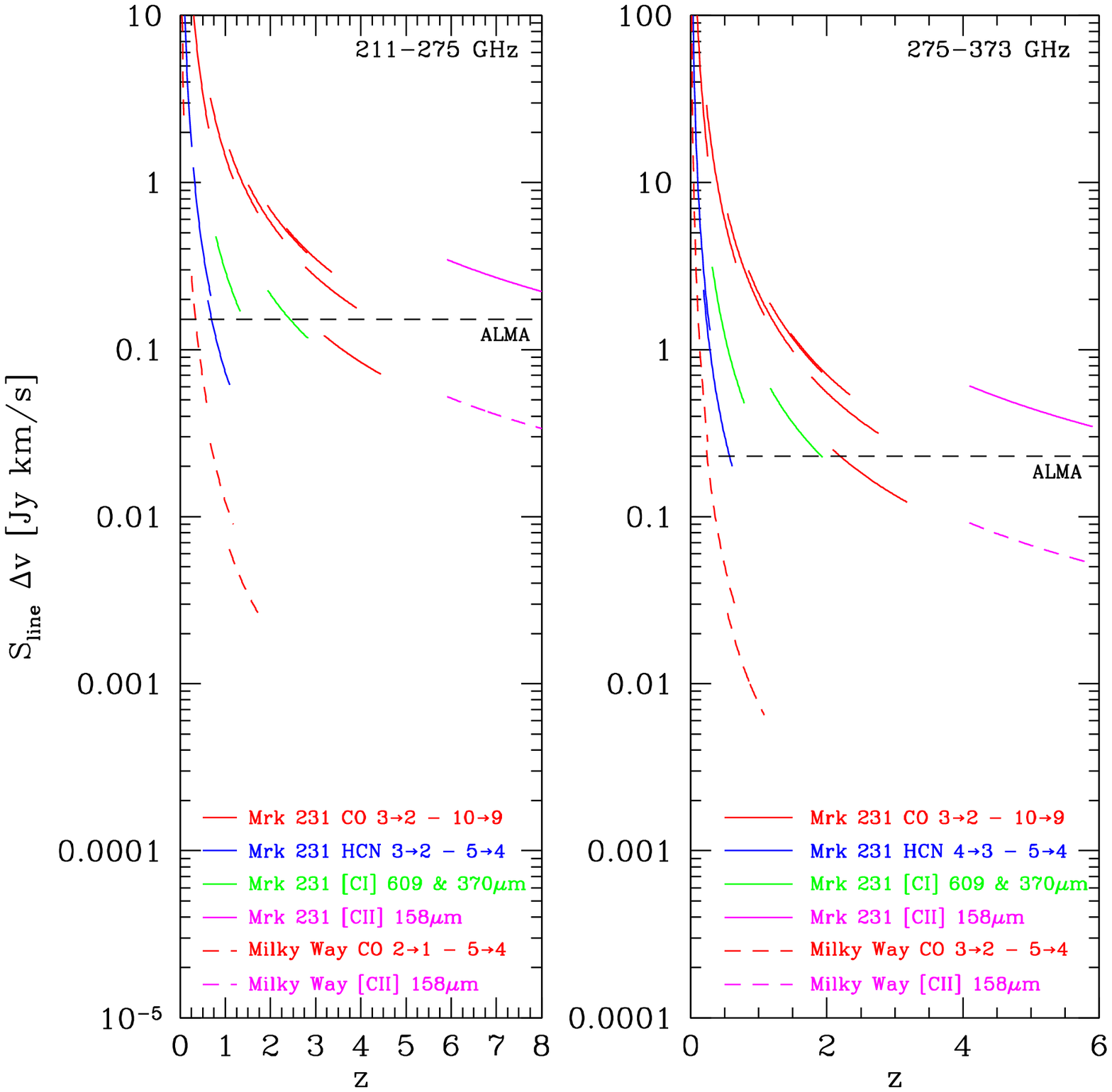}
\caption{(continued)}
\end{figure}

\clearpage

\begin{figure}
\plotone{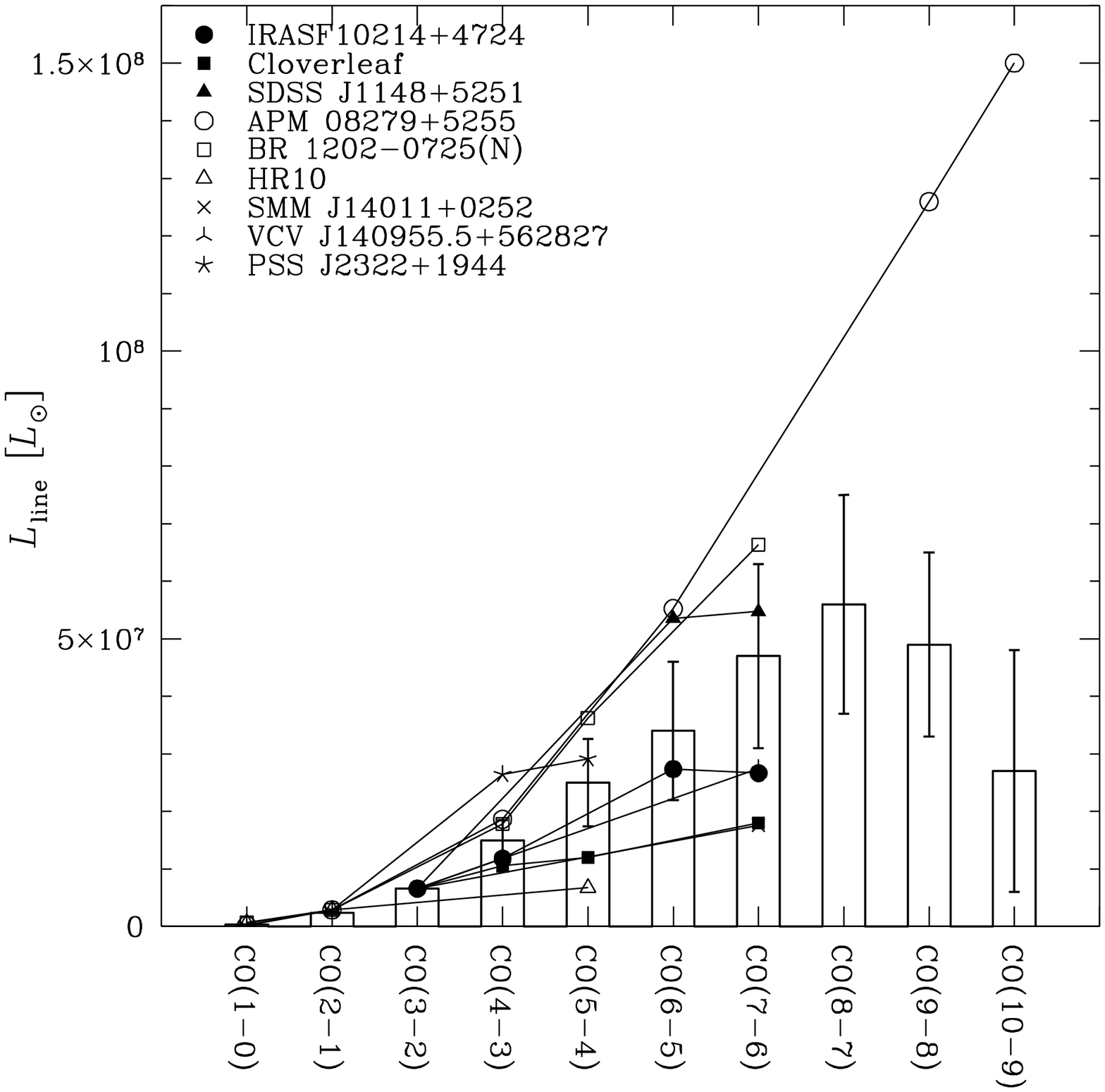}
\caption{CO  line   luminosities  in  various   high-redshift  objects
  compared to the Mrk\,231 template. Lines luminosities are normalized
  to   the  Mrk\,231   values   at  either   CO   $J{=}3{-}2$  or   CO
  $J{=}2{-}1$. Observed line fluxes  are taken from the compilation by
  Solomon \&  Vanden Bout (2005),  supplemented with more  recent data
  from Riechers et al.\ (2006).}
\end{figure}

\clearpage

\begin{deluxetable}{lcc}
\tablecolumns{3}
\tablewidth{0pc}
\tablecaption{CO J=1--0 fluxes measured for Mrk\,231}
\tablehead{
\colhead{Telescope}&\colhead{$\rm S_{line}$ (Jy\,km\,s$ ^{-1}$)}&\colhead{Reference}}   
\startdata
 IRAM 30-meter       &   $100\pm 20$ & Solomon et al. 1997 \\
 NRAO 12-meter       &   $100\pm 20$ & Papadopoulos \& Seaquist 1998\\
 FCRAO               &   $82\pm 15$  & Young et al. 1995\\
 IRAM 30-meter       &   $72\pm 15$  & Kr\"ugel et al. 1990\\
 OVRO                &   $62\pm 10$  & Bryant \& Scoville 1996\\
 IRAM PdBI           &   $68\pm 10$  & Downes \& Solomon 1998\\
\enddata
\end{deluxetable}

\clearpage

\begin{deluxetable}{lccc}
\tablecolumns{8}
\tablewidth{0pc}
\tablecaption{CO and HCN line fluxes and ratios}
\tablehead{
\colhead{Transition}&\colhead{$\rm S_{line}$}&\colhead{Line ratio\tablenotemark{a}}
 &\colhead{References\tablenotemark{c}}\\
                    & (Jy\,km\,s$ ^{-1}$)    &  & }   
\startdata
 CO J=1--0           &   $88\pm 9$\tablenotemark{b}& \nodata    & Table 1 (single dish data) \\
 CO J=2--1           &   $315\pm 30$\tablenotemark{b}            &   $0.90\pm 0.13$     & 1, 2, 3\\
 CO J=3--2           &   $568\pm 80$\tablenotemark{b}            &   $0.71\pm 0.12$     & 4, this work\\
 CO J=4--3           &   $980\pm 230$           &   $0.70\pm 0.18$     & this work\\
 CO J=6--5           &   $1465\pm 500$          &   $0.46\pm 0.16$     & this work\\
\hline
 HCN J=1--0          &   $15\pm 3$              &   \nodata           & 5\\
 HCN J=4--3          &   $65\pm 13$             &   $0.27\pm 0.08$     & this work\\
\enddata
\tablenotetext{a}{Velocity/area-averaged brightness temperature line \\
\hspace*{0.45cm} ratios $\rm r_{J+1\,J}=\langle T_b(J+1,J)\rangle/\langle 
T_b(1,0)\rangle$.}
\tablenotetext{b}{Averages of the values extracted from the data in the\\
\hspace*{0.45cm} listed references.}
\tablenotetext{c}{1. Papadopoulos \& Seaquist 1998; 2. Downes \& Solomon 1998~(two \\
\hspace*{0.45cm} values reported); 3. Glenn, \& Hunter 2001 (CSO for $\rm S/T\sim 
\hspace*{0.45cm} 50\, Jy/K$); 4. Lisenfeld et al. 1996;  5. Solomon, Downes, \& Radford\\
\hspace*{0.45cm} 1992a}
\end{deluxetable}

\clearpage

\begin{deluxetable}{ccccc}
\tablecolumns{5}
\tablewidth{0pc}
\tablecaption{Physical conditions of the dense gas phase}
\tablehead{
\colhead{\rm $\rm T_k$\tablenotemark{a}} &\colhead{$\rm n(H_2)$\tablenotemark{a}} &
 \colhead{$\rm \Lambda _{HCN}$\tablenotemark{a}} &
\colhead{$\rm K_{vir}$\tablenotemark{b}} & \colhead{$\rm R_{HCN/CO}\,(r_{43}(HCN))$\tablenotemark{c}}  \\
       (K)           &  ($\rm cm^{-3}$)  & $\rm (km\, s^{-1}\, pc^{-1})^{-1}$ &  & }
 \startdata
15     &  $3\times 10^5$ & $\rm 3\times 10^{-10}$  & $5.9\alpha^{-1/2}$  & 2.67 (0.26) \\
20-25  &  $3\times 10^4$ & $3\times 10^{-8}$       & $0.18\alpha^{-1/2}$ & 1.97-1.55 (0.24-0.28)\\
30     &  $3\times 10^5$ & $ 10^{-10} $            & $17.8\alpha^{-1/2}$ & 0.90 (0.27)\\
35     &  $10^5$         & $ 10^{-9}  $            & $3.08\alpha^{-1/2}$ & 0.80 (0.28)\\
{\bf 40-45}\tablenotemark{d} & ${\bf 3\times 10^4}$ & ${\bf 10^{-8}}$ & ${\bf 0.56\alpha^{-1/2}}$ & {\bf 0.72-0.64 (0.28-0.30)}\\
{\bf 50-70}\tablenotemark{d} & ${\bf 10^4}$ & ${\bf 10^{-7}}$       & ${\bf 0.1\alpha^{-1/2}}$ & {\bf 0.65-0.47 (0.23-0.28)}\\
75-90  &  $10^5 $        & $3\times 10^{-10}$      & $10.25\alpha^{-1/2}$& 0.25-0.21 (0.25-0.27)\\
95-105 &  $3\times 10^4$ & $3\times 10^{-9}$       & $1.87\alpha^{-1/2}$ & 0.21-0.19 (0.27-0.28)\\
110-150&  $10^4$         & $3\times 10^{-8}$       & $0.32\alpha^{-1/2}$ & 0.19-0.15 (0.26-0.30)\\
\enddata
\tablenotetext{a}{Parameters corresponding to best LVG solutions for $\rm r_{43}(HCN)$ ratio\\
\hspace*{0.45cm} ($\chi ^2\la 0.3$; see section 3)}
\tablenotetext{b}{From Equation~2,  $\rm r_{HCN}=[HCN/H_2]=2\times 10^{-8}$  (Irvine, Goldsmith \& \\
\hspace*{0.45cm} Hjalmarson 1987; Lahuis \& van Dishoeck 2000) ($\alpha \sim 1-2.5$ see text).}
\tablenotetext{c}{The (dense gas)-dominated HCN(1--0)/CO(6--5) and HCN(4--3)/(1--0)\\
\hspace*{0.45cm}  brightness temperature line ratios for each LVG set of parameters.\\
\hspace*{0.45cm}  Observed ratios: $\rm R_{HCN/CO}=0.62\pm0.24$ and  $\rm r_{43}(HCN)=0.27\pm 0.08$).}
\tablenotetext{d}{The best solution ranges are indicated in boldface.}

\end{deluxetable}
\clearpage

\begin{deluxetable}{ccccc}
\tablecolumns{5}
\tablewidth{0pc}
\tablecaption{Cooling budget of the molecular gas in Mrk\,231}
\tablehead{
\colhead{\rm Transition} & \colhead{$S_{\rm line}$} & 
\colhead{$L'_{\rm line}$} & \colhead{$L_{\rm line}$} & \colhead{Notes} \\
  &  [${\rm Jy\,km\,s^{-1}}$]  & [${\rm K\,km\,s^{-1}\,pc^2}$] 
  & [$L_{\odot}$] & }
 \startdata
CO $J=1-0$ &  88 &  $6.9\cdot10^9$ & $3.4\cdot10^5$ & \\
CO $J=2-1$ &  315 & $6.2\cdot10^9$ & $2.4\cdot10^6$ & \\
CO $J=3-2$ &  568 & $5.0\cdot10^9$ & $6.6\cdot10^6$ & \\
CO $J=4-3$ &  980 & $4.8\cdot10^9$ & $1.5\cdot10^7$ & \\
CO $J=5-4$ &      & $4.0\cdot10^9$ & $2.5\cdot10^7$ & \tablenotemark{a} \\
CO $J=6-5$ & 1465 & $3.2\cdot10^9$ & $3.4\cdot10^7$ & \\
CO $J=7-6$ &      & $2.8\cdot10^9$ & $4.7\cdot10^7$ & \tablenotemark{b} \\
CO $J=8-7$ &      & $2.2\cdot10^9$ & $5.6\cdot10^7$ & \tablenotemark{b} \\
CO $J=9-8$ &      & $1.4\cdot10^9$ & $4.9\cdot10^7$ & \tablenotemark{b} \\
CO $J=10-9$ &     & $5.5\cdot10^8$ & $2.7\cdot10^7$ & \tablenotemark{b} \\
CO dense phase   & & & $2.5\cdot10^8$ & \tablenotemark{c} \\
CO diffuse phase & & & $8.9\cdot10^6$ & \tablenotemark{d} \\
\hbox{[\ci]} $609\mum$ & 200 & $8.7\cdot10^8$ & $3.3\cdot10^6$ & \\
\hbox{[\ci]} $390\mum$ &     & $8.7\cdot10^8$ & $1.5\cdot10^7$ & \tablenotemark{e} \\
\hbox{[\cii]} $158\mum$ &     &       & $3.4\cdot10^8$ & \\
\enddata
\tablenotetext{a}{~line strength estimated by interpolation (CO $\rm J+1\rightarrow J$, $\rm J+1>4$\\
\hspace*{0.45cm} lines are dominated  by the dense gas phase).}
\tablenotetext{b}{~line strength estimated using the mean value resulting from the\\
\hspace*{0.45cm} LVG models}
\tablenotetext{c}{sum of CO line strengths from the dense phase using the mean\\
\hspace*{0.45cm} of our LVG models}
\tablenotetext{d}{sum of observed CO line strengths minus modeled CO line\\
 \hspace*{0.45cm} strengths from the dense phase}
\tablenotetext{e}{~line strength estimated from the $609\mum$ line (see text)}
\end{deluxetable}

\clearpage

\begin{deluxetable}{ccccc}
\tablecolumns{5}
\tablewidth{0pc}
\tablecaption{Observed and modeled HCN line luminosities of Mrk\,231}
\tablehead{
\colhead{\rm Transition} & \colhead{$S_{\rm line}$} & 
\colhead{$L'_{\rm line}$} & \colhead{$L_{\rm line}$} & \colhead{Notes} \\
  &  [${\rm Jy\,km\,s^{-1}}$]  & [${\rm K\,km\,s^{-1}\,pc^2}$] 
  & [$L_{\odot}$] & }
 \startdata
HCN $J=1-0$ & 15 & $2.0\cdot10^9$ & $4.5\cdot10^4$ & \\
HCN $J=2-1$ &    & $1.6\cdot10^9$ & $2.9\cdot10^5$ & \tablenotemark{a} \\
HCN $J=3-2$ &    & $9.7\cdot10^8$ & $5.8\cdot10^5$ & \tablenotemark{a} \\
HCN $J=4-3$ & 65 & $5.5\cdot10^8$ & $7.7\cdot10^5$ & \\
HCN $J=5-4$ &    & $2.7\cdot10^8$ & $7.4\cdot10^5$ & \tablenotemark{a} \\
HCN $J=6-5$ &    & $9.2\cdot10^7$ & $4.4\cdot10^5$ & \tablenotemark{a} \\
HCN $J=7-6$ &    & $2.5\cdot10^7$ & $1.9\cdot10^5$ & \tablenotemark{a} \\
\enddata
\tablenotetext{a}{~line strength estimated using the mean value resulting from\\
\hspace*{0.45cm} the two best ranges of LVG solutions (Table 3)}
\end{deluxetable}

\clearpage

\begin{deluxetable}{cccc}
\tablecolumns{5}
\tablewidth{0pc}
\tablecaption{Conditions compatible with the CO (6--5)/(4--3) ratio\tablenotemark{a}}
\tablehead{
\colhead{\rm $\rm T_k$} &\colhead{$\rm n(H_2)$} & \colhead{$\rm \Lambda _{CO}$} &
\colhead{$\rm K_{vir}$\tablenotemark{b} ($\tau _{10}$\tablenotemark{c} )}\\
(K) & ($\rm cm^{-3}$)  & $\rm (km\, s^{-1}\, pc^{-1})^{-1}$ &  }
\startdata
15              & $3\times 10^5$ &  $3\times 10^{-4}$& $0.03\alpha ^{-1/2}$ ($\gg 1$)\\
20              & $10^4$         &  $3\times 10^{-4}$& $0.16\alpha ^{-1/2}$ ($\gg 1$)\\
25              & $10^4$         &  $10^{-4}$        & $0.48\alpha ^{-1/2}$ ($\gg 1$)\\
30-35           & $3\times 10^3$ &  $3\times 10^{-4}$& $0.3\alpha ^{-1/2}$  ($\gg 1$)\\
40-45           & $3\times 10^4$ &  $10^{-6}$        & $28\alpha ^{-1/2}$   (1.2-1.5)\\
50-75           & $3\times 10^3$ &  $10^{-4}$        & $0.89\alpha ^{-1/2}$ (5-10)\\
80-135          & $10^4$         &$3\times 10^{-6}$  &$16\alpha ^{-1/2}$    (0.08-0.45)\\
140-150         & $3\times 10^3$ & $3\times 10^{-5}$ & $2.96\alpha ^{-1/2}$ (0.77-0.88)\\
\enddata
\tablenotetext{a}{Observed value: $\rm r_{65/43}=0.66\pm0.26$}
\tablenotetext{b}{From Equation~2, and $\rm r_{CO}=[CO/H_2]=10^{-4}$.}
\tablenotetext{c}{The optical depth of the CO J=1--0 line.}
\end{deluxetable}

\end{document}